\documentclass[preprint,12pt]{elsarticle}
\usepackage{graphics}
\usepackage{subfigure}
\usepackage{graphicx}
\usepackage{amsmath}
\usepackage{float}
\usepackage{amssymb}
\usepackage{bm}

\journal{PhysicaA}
\begin{document}
\begin{frontmatter}



\title{Transport of Finite Size Self-Propelled Particles Confined in a $2D$ Zigzag Channel with Gaussian Colored Noise}

\author{Bing Wang}
{\ead{hnitwb@163.com}}
\author{Yong Wu}
\author{Xiaoxiao Zhang}
\author{Hao Chen}
\address{School of Mechanics and Optoelectronics Physics, Anhui University of Science and Technology, Huainan, 232001, P.R.China}

\begin{abstract}
The directional transport of finite size self-propelled Brownian particles confined in a $2D$ zigzag channel with colored noise is investigated. The noises(noise parallel to $x$-axis and $y$-axis), the asymmetry parameter $\Delta k$, the ratio $f$(ratio of the particle radius and the bottleneck half width), the self-propelled speed $v_0$ have joint effect on the particles. The average velocity of self-propelled particles is significantly different from passive particles. The average velocity exhibits complicated behavior with increasing self-propelled speed $v_0$.
\end{abstract}

\end{frontmatter}

\section{\label{label1}Introduction}
The transport properties of Brownian particles confined in $2D$ or $3D$ channel is a key issue for a variety of situations in recent years due to its ubiquitous importance in many disciplinaries ranging from physicochemical to biological systems. Some biological processes such as ion pumping, neuronal signaling, porous media, and photosynthesis, rely on the transport of ions across membranes or through channels\cite{Borromeo,Costantini,Dan,Heinsalu2004,Tatarkova,Lindenberg,Reimann,Khoury,Reguera2006,Machura,Tierno,Heinsalu2008,Coffey,Mei,Simon,Li,Liu2017}. Brownian particles in regular arrays of rigid obstacles, and also in the corrugated geometry channel show many interesting phenomena.

Self-propelled particles can perform active Brownian motion by extracting energy from external environment. Self-propelled particles confined in channel has attracted widely attention and shown lots of interesting phenomenon. Malgaretti \emph{et al}. analyzed the dynamics of Brownian ratchets in a confined environment and found the combined rectification mechanisms may lead to bidirectional transport\cite{Malgaretti}. Teeffelen \emph{et al}. studied the motion of a chiral swimmer in a confining channel and found self-propelled particles move along circles rather than along a straight line when their driving force does not coincide with their propagation direction\cite{Teeffelen}. Pototsky \emph{et al}. considered a colony of point like self-propelled particles without direct interactions that cover a thin liquid layer on a solid support\cite{Pototsky}. Wu \emph{et al}. investigated the rectification transport of finite finite size self-propelled particles in a two dimensional asymmetric channel and found average velocity in the presence of translational noise may be orders of magnitude larger than that in the absence of translational noise\cite{Wu}. Ao \emph{et al}. investigated the transport diffusivity of Janus particles in the absence of external biases, and found the self-diffusion constants depends on both the strength and the chirality of the self-propulsion mechanism, and self-diffusion can be controlled by tailoring the compartment geometry in a periodic channel\cite{Ao}. Liu \emph{et al}. investigated the entropic stochastic resonance phenomenon when a self-propelled Janus particle moves in a double-cavity container and found the entropic stochastic resonance can survive even if there is no symmetry breaking in any direction\cite{Liu}.  Liao \emph{et al}. investigated transport and diffusion of paramagnetic ellipsoidal particles under the action of a rotating magnetic field in a two-dimensional channel\cite{Liao}.

In this paper, we investigate the transport phenomenon of finite size self-propelled Brownian particles confined in $2D$ channel with colored noise. The paper is organized as follows: In Section \ref{label2}, the basic model is provided. In Section \ref{label3}, the effects of parameters is investigated by means of simulations. In Section \ref{label4}, we get the conclusions.

\section{\label{label2}Basic model and methods}
In the present work, we consider the self-propelled Brownian particles confined in a $2D$ zigzag channel. The dynamics of the particles can be described by the following Langevin equations\cite{Wu, Reguera2012}
\begin{equation}
\gamma\frac{d\bm{r}}{dt}=\frac{v_0}{\mu}\bm{n}+\sqrt{\gamma}\bm{\xi}(t), \label{Ert}
\end{equation}
\begin{equation}
\gamma_{\theta}\frac{d\theta}{dt}=\sqrt{\gamma_{\theta}}\xi_{\theta}({t}). \label{Ethetat}
\end{equation}
where $\bm{r}(x,y)$ is the position of the particle center of mass. $v_0$ is the self-propelled speed. $\bm{n}=(\cos\theta,\sin\theta)$ is the unit vector. $v_0$ is the self-propelled speed, $\mu$ is the mobility. $\theta$ is the self-propelled angle, and denotes its direction with respect to the channel axis. $\bm{\xi}(t)=(\xi_{{x}}(t),\xi_{{y}}(t))$ is the Gaussian colored noise. $\gamma=6\pi vR$($R$ is the particle radius, $v$ is the shear viscosity of the fluid) is the friction coefficient and satisfies the Stokes law. $\gamma_{\theta}$ is the rotational friction coefficient.

\begin{figure}
\centering
\subfigure{
\includegraphics[height=6cm,width=6cm]{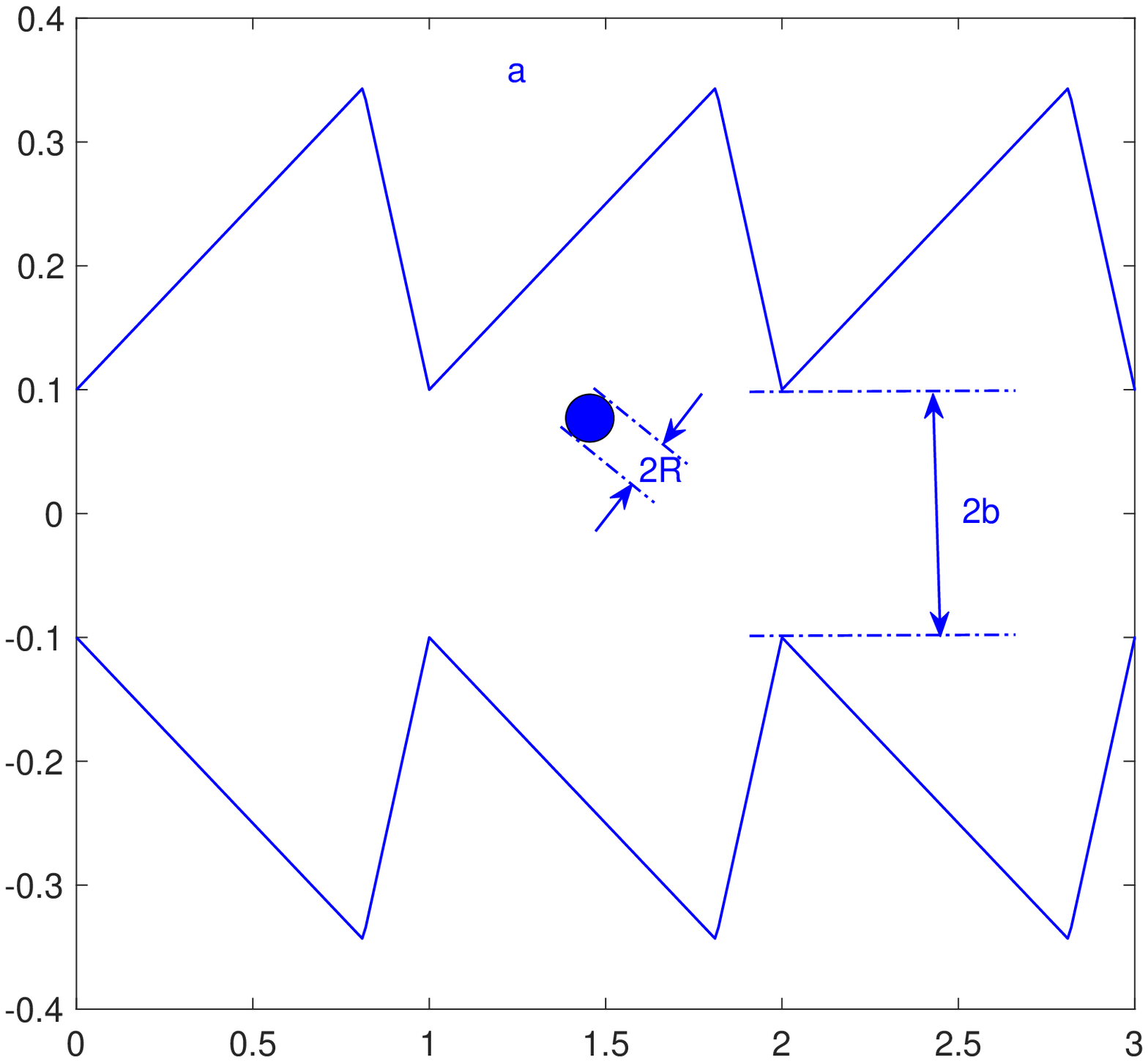}
}
\subfigure{
\includegraphics[height=6cm,width=6cm]{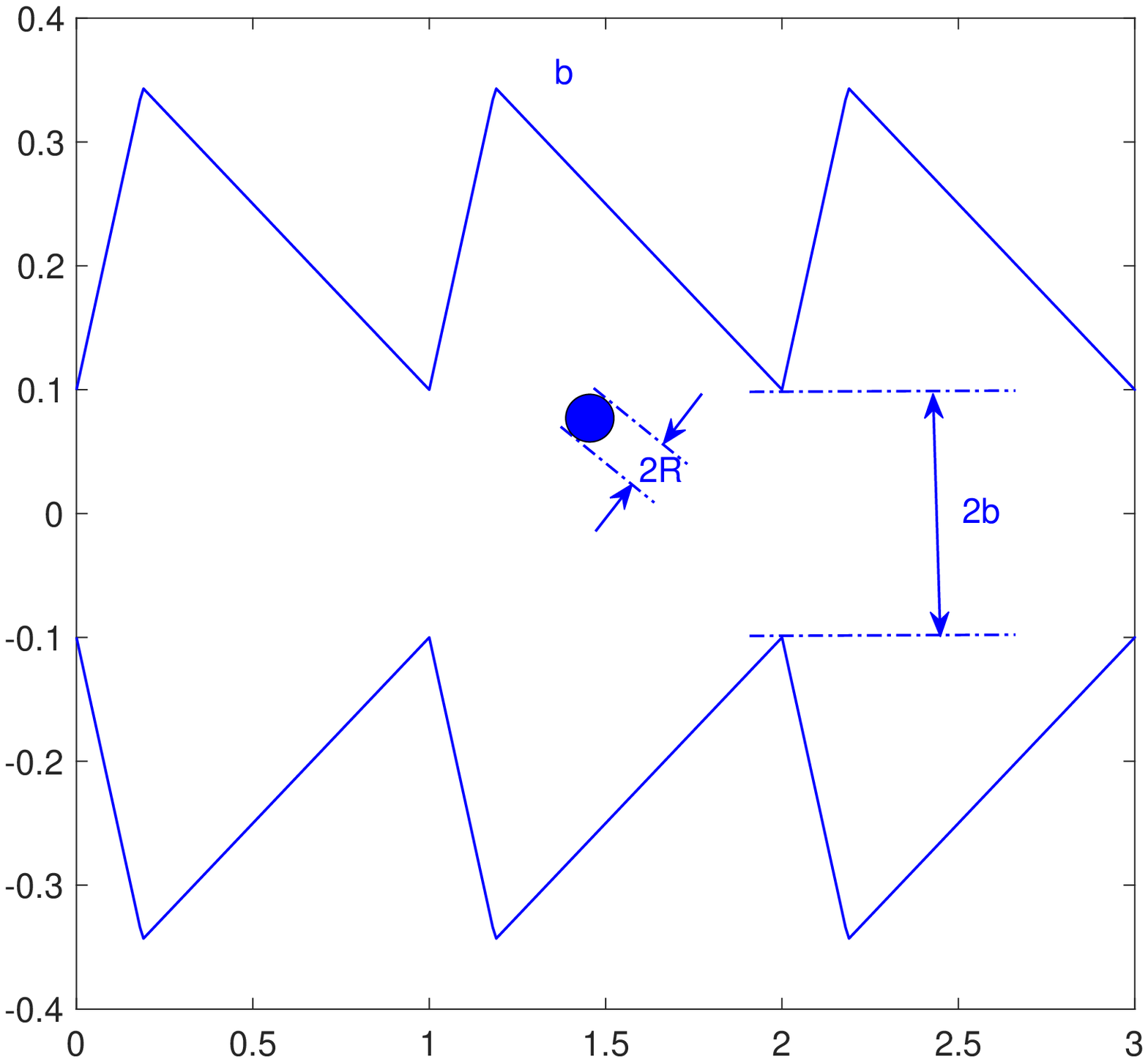}
}
\caption{Schematic of $2D$ channel with periodicity $L$, $R$ is the radius of the particle with $L=1.0$, $b=0.2$, $\bar{k}=0.8$:(a)$\Delta k=-0.5$, (b)$\Delta k=0.5$.}\label{Channel}
\end{figure}

The confined zigzag channel is a periodic function in space along the $x$ direction with period $L$(depicted in Fig.\ref{Channel}). The walls of the cavity have been modelled by the following  piecewise function
\begin{equation}
w({x})=\left\{
\begin{array}{rcl}
b+k_1\bar{x} & & {\bar{x}<L_0}\\
b+k_2(L-\bar{x}) & & {\bar{x}\geq L_0}\\
\end{array}
\right.
\end{equation}

The upper and lower boundary functions are $w(x)$ and $-w(x)$, respectively. $b$ is the half width of the bottleneck. $k_1$ and $k_2$ are the slopes of the walls. $L_0=Lk_2/(k_1+k_2)$ indicates the location of the point of maximum width, and $\bar{x}=x$ mod $L$ is the modulo function. For convenience, we define $\bar{k}=(k_1+k_2)/2$ and $\Delta {k}=(k_1-k_2)/2$. $\Delta k$ reflects the asymmetry of the channel. The channel is symmetric at $\Delta {k}=0$ and straight at $\bar{k}=\Delta {k}$.

For a hard finite size real particle with the radius $R$ in the channel, the available space from the walls is described by the following function

\begin{equation}
w_u({x})=\left\{
\begin{array}{rcl}
b-\sqrt{R^2-\bar{x}^2} & & {0\leq \bar{x}<L_a}\\
b+k_1\bar{x}-R\sqrt{1+k_1^2} & & {L_a\leq \bar{x}<L_b}\\
b+k_2(L-\bar{x})-R\sqrt{1+k_2^2} & & {L_b\leq \bar{x}<L_c}\\
b-\sqrt{R^2-(\bar{x}-L)^2} & & {L_c\leq \bar{x}<L}
\end{array}
\right.
\end{equation}

$L_a=Rk_1/\sqrt{1+k_1^2}$, $L_b=C+R(\sqrt{1+k_1^2}-\sqrt{1+k_1^2})k_1/(k_1+k_2)$ and  $L_c=L-Rk_2/\sqrt{1+k_1^2}$

Upon introducing characteristic length scale $L$, time scale $L^2 6\pi vb$. Eqs. (\ref{Ert}, \ref{Ethetat}) can be rewritten in dimensionless form

\begin{equation}
\frac{d\bm{\hat{r}}}{d\hat{t}}=\frac{\hat{v}_0}{f}\bm{n}+{\frac{1}{\sqrt f}}\bm{\xi}(\hat{t}), \label{Ehatrt}
\end{equation}
\begin{equation}
\frac{d\theta}{d\hat{t}}=\xi_{\theta}({\hat{t}}). \label{Ehatthetat}
\end{equation}
here, $\bm{\hat{r}}=\bm{{r}}/L$, $\hat{t}=t/\tau$, and $\tau=L^2 6\pi vb$, $\hat{v}_0=v_0L/{\mu}$. $f=R/b$ is the ratio of the particle radius $R$ and the bottleneck half width $b$. Because the size of the particle is finite, so $R>0$, and the ratio $f>0$ too. On the other hand, the particle mast be able to past through the bottleneck of the channel, so $R\leq b$, and $f\leq1$.  In a word, $0<f\leq1$.

$\xi_x$ and $\xi_y$ are Gaussian colored noises. $\xi_x$ parallel to $x$-axis, and $\xi_y$ parallel to $y$-axis, respectively. $\xi_{\theta}$ is the self-propelled angle Gaussian colored noise, and describes the nonequilibrium angular fluctuation. $\xi_{x}$, $\xi_{y}$ and $\xi_{\theta}$ satisfy the following relations
\begin{equation}
\langle\xi_i(t)\rangle=0,(i=x,y,\theta)
\end{equation}
\begin{equation}
\langle\xi_i(t)\xi_j(t')\rangle=\delta_{ij}\frac{Q_i}{\tau_i}\exp[-\frac{|t-t'|}{\tau_i}],(i=x,y,\theta)
\end{equation}
$\langle\cdots\rangle$ denotes an ensemble average over the distribution of the random forces. $Q_i(i=x,y,\theta)$ is the noise intensity of $\xi_i(i=x,y,\theta)$. $\tau_i(i=x,y,\theta)$ is the self-correlation time of $\xi_i(i=x,y,\theta)$.

In the following, we will only use dimensionless variables for simplicity and shall omit the hat notation for all quantities.

A central practical question in the theory of Brownian motors is the over all long time behavior of the particle, and the key quantities of particle transport is the particle velocity $\langle V\rangle$. Because particles along the $y$ direction are confined, we only calculate the $x$ direction average velocity $\langle V\rangle$ based on Eqs.(\ref{Ehatrt},\ref{Ehatthetat}).
\begin{equation}
\langle V\rangle=\lim_{t\to\infty}\frac{\langle{x(t)-x(t_0)}\rangle}{t-t_0},
\end{equation}
$x(t_0)$ is the position of particles at time $t_0$.

\section{\label{label3}Results and discussion}
In order to give a simple and clear analysis of the system. Eqs.(\ref{Ehatrt}) and (\ref{Ehatthetat}) are integrated using the Euler algorithm. The total integration time was more than $10^5$ and the integration step time $\Delta t=10^{-4}$. The stochastic averages were obtained as ensemble averages over $10^5$ trajectories. With these parameters, the simulation results do not depend on the time step, the integration time, and the number of trajectories.

\begin{figure}
\centering
\subfigure{
\includegraphics[height=6cm,width=6cm]{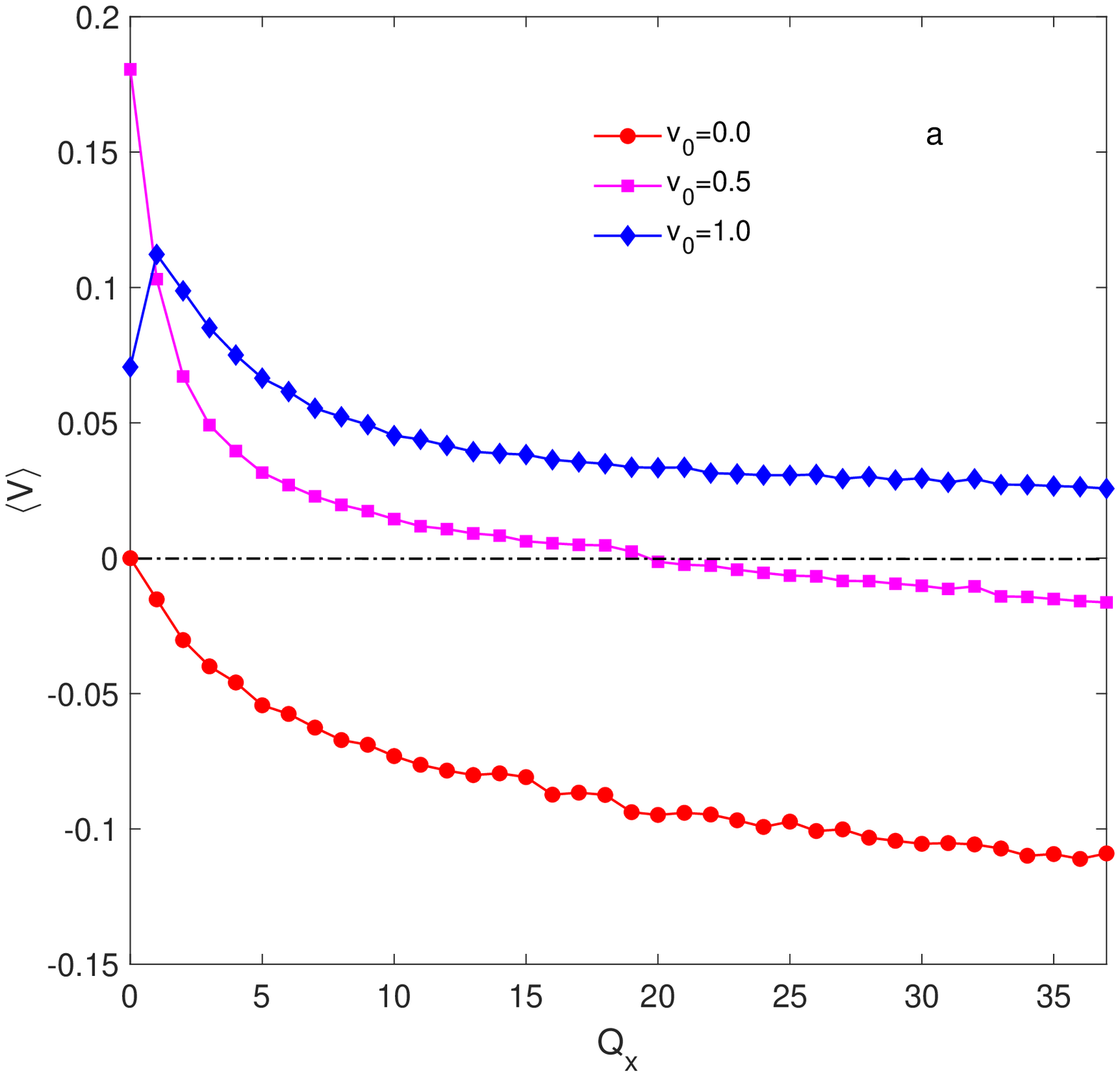}
}
\subfigure{
\includegraphics[height=6cm,width=6cm]{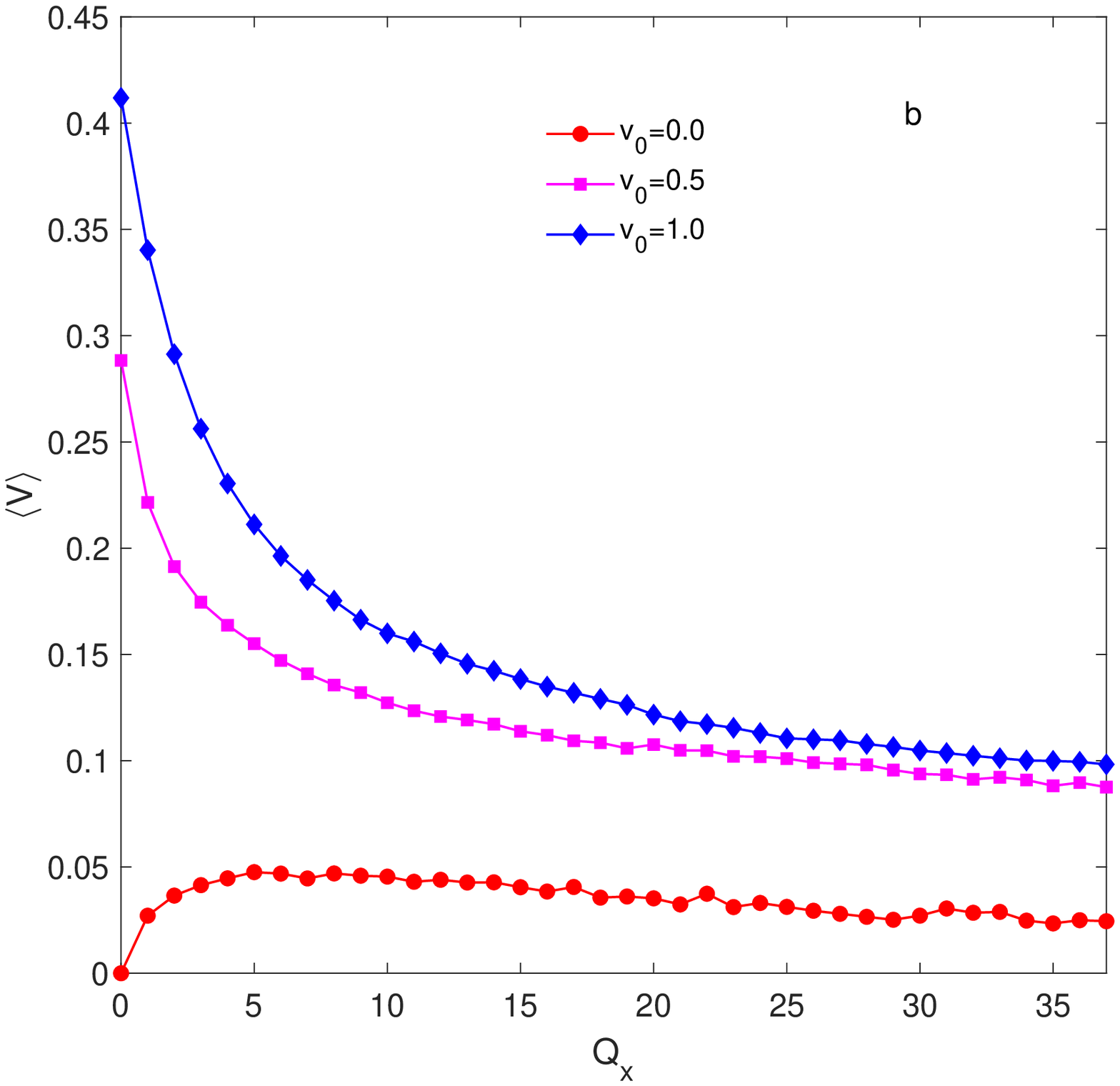}
}
\caption{The average velocity $\langle V\rangle$ as a function of $x$-axis noise intensity $Q_x$ with different $v_0$. The other parameters are $Q_y=2.0$, $Q_{\theta}=0.2$, $\tau_x=\tau_y=\tau_{\theta}=1.0$, $L=1.0$, $\bar{k}=1.0$, $f=0.5$:(a)$\Delta k=-0.5$, (b)$\Delta k=0.5$.}
\label{VQx}
\end{figure}

The average velocity $\langle V\rangle$ as a function of the $x$-axis noise intensity $Q_x$ with different self-propelled speed $v_0$ is reported in Fig.\ref{VQx}. In Fig.\ref{VQx}(a)($\Delta k=-0.5$), we find the average velocity $\langle V\rangle$ decreases monotonically with increasing $Q_x$ when $v_0=0$ and $v_0=0.5$, and $\langle V\rangle$ has a maximum with increasing $Q_x$ when $v_0=1.0$. The particle is passive as $v_0=0$, so passive particle moves in $-x$ direction, and the moving speed(in $-x$ direction) increases with increasing $Q_x$. In Fig.\ref{VQx}(a), $v_0=0.5$, $\langle V\rangle>0$ when $Q_x <20$, and $\langle V\rangle<0$ when $Q_x >20$, so the transport reverse phenomenon appears with increasing $x$-axis noise intensity $Q_x$. In Fig.\ref{VQx}(b)($\Delta k=0.5$), we find the average velocity $\langle V\rangle >0$ when $v_0=0$($v_0=0.5$ and $v_0=1.0$), so the particle moves in $+x$ direction when $\Delta k=0.5$. $\langle V\rangle$ decreases with increasing $Q_x$ when $v_0=0.5$ and $v_0=1.0$, large $x$-axis noise intensity will weaken $+x$ directional movement. $\langle V\rangle$ has a maximum with increasing $Q_x$ when the particle is passive($v_0=0.0$).

\begin{figure}
\centering
\subfigure{
\includegraphics[height=6cm,width=6cm]{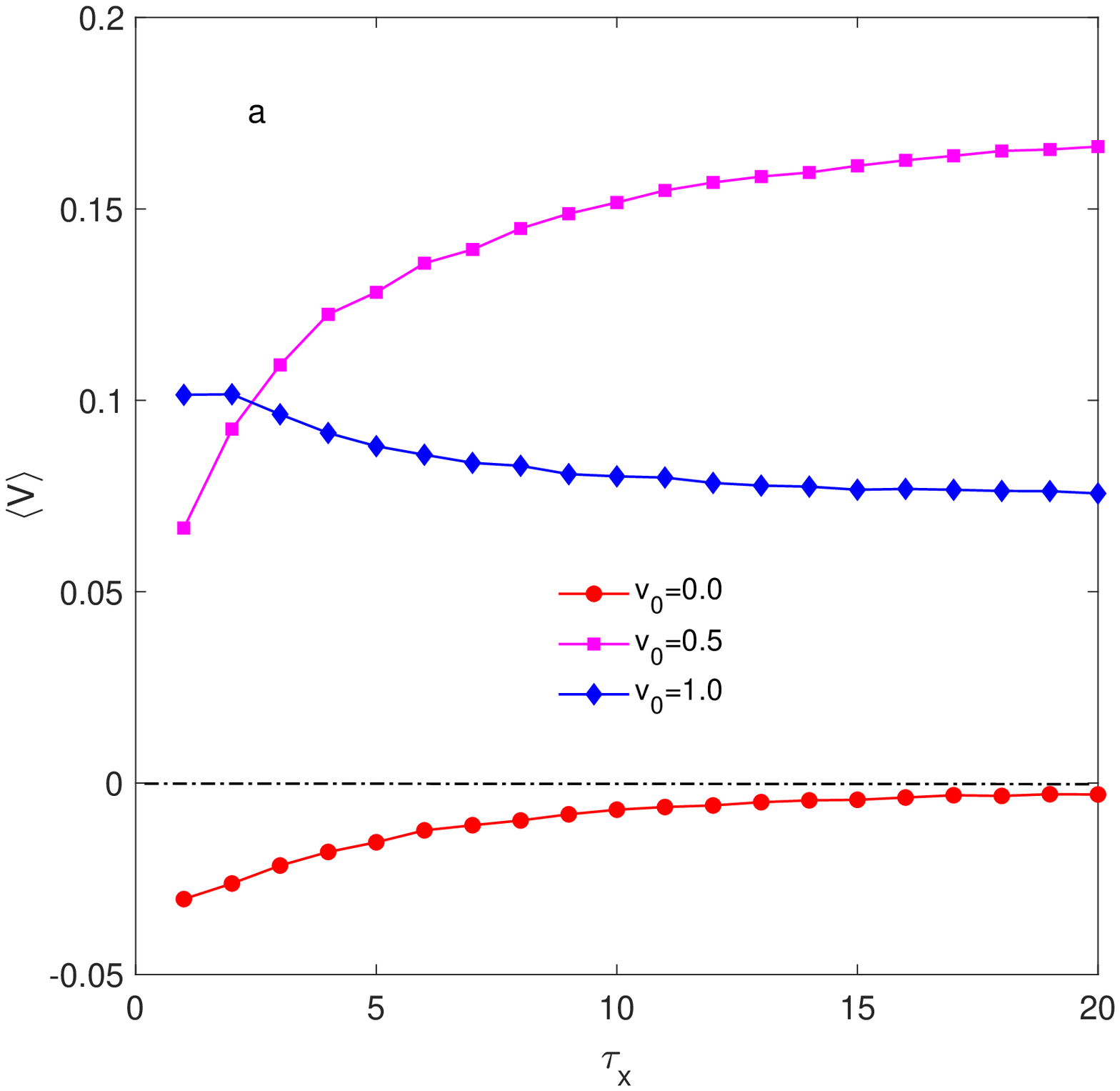}
}
\subfigure{
\includegraphics[height=6cm,width=6cm]{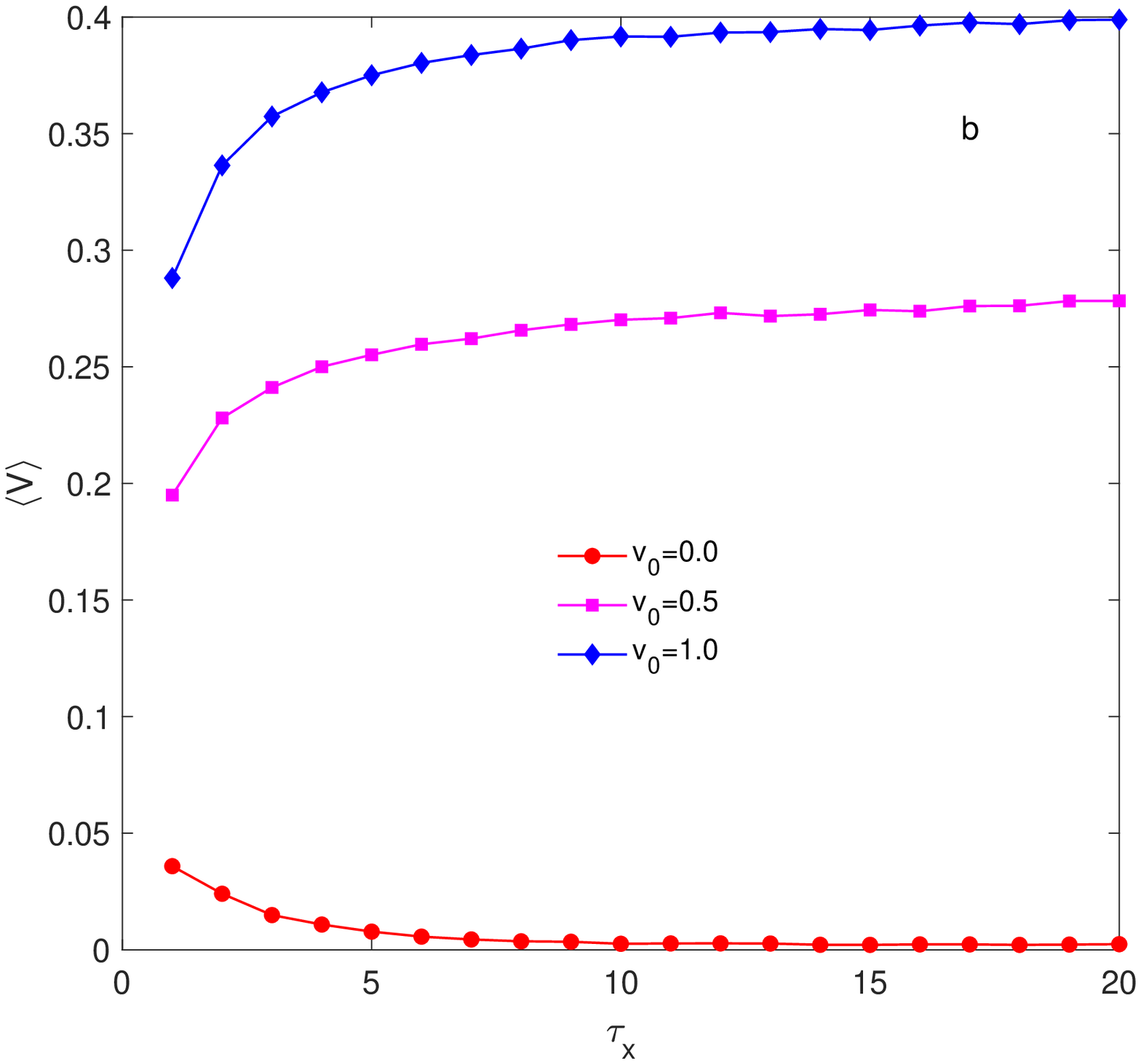}
}
\caption{The average velocity $\langle V\rangle$ as a function of $x$-axis noise self-correlation time $\tau_x$ with different $v_0$. The other parameters are $Q_x=Q_y=2.0$, $Q_{\theta}=0.2$, $\tau_y=\tau_{\theta}=1.0$, $L=1.0$, $\bar{k}=1.0$, $f=0.5$:(a)$\Delta k=-0.5$, (b)$\Delta k=0.5$.}
\label{Vtaux}
\end{figure}

The average velocity $\langle V\rangle$ as a function of the $x$-axis noise self-correlation time $\tau_x$ with different $v_0$ is reported in Fig.\ref{Vtaux}. In Fig.\ref{Vtaux}(a)($\Delta k=-0.5$), we find $\langle V\rangle$ increases monotonically with increasing $\tau_x$ when $v_0=0$ and $v_0=0.5$. $\langle V\rangle<0$ when $v_0=0.0$, so passive particle($v_0=0$) moves in $-x$ direction, and the moving speed(in $-x$ direction) decreases with increasing $\tau_x$. In Fig.\ref{Vtaux}(a), we can also find $\langle V\rangle$ decreases with increasing $\tau_x$ when $v_0=1.0$. In Fig.\ref{Vtaux}(b)($\Delta k=0.5$), $\langle V\rangle>0$ when $v_0=0.0$($v_0=0.5$ and $v_0=1.0$), so the particle moves in $+x$ direction. $\langle V\rangle$ increases with increasing $\tau_x$ when $v_0=0.5$ and $v_0=1.0$, but there is almost no change for $\langle V\rangle$ with increasing $\tau_0$ when $\tau_0$ is large. $\langle V\rangle$ decreases with increasing $\tau_x$ when $v_0=0.0$, and $\langle V\rangle$ shows little change when $\tau_0$ is large.

\begin{figure}
\centering
\subfigure{
\includegraphics[height=6cm,width=6cm]{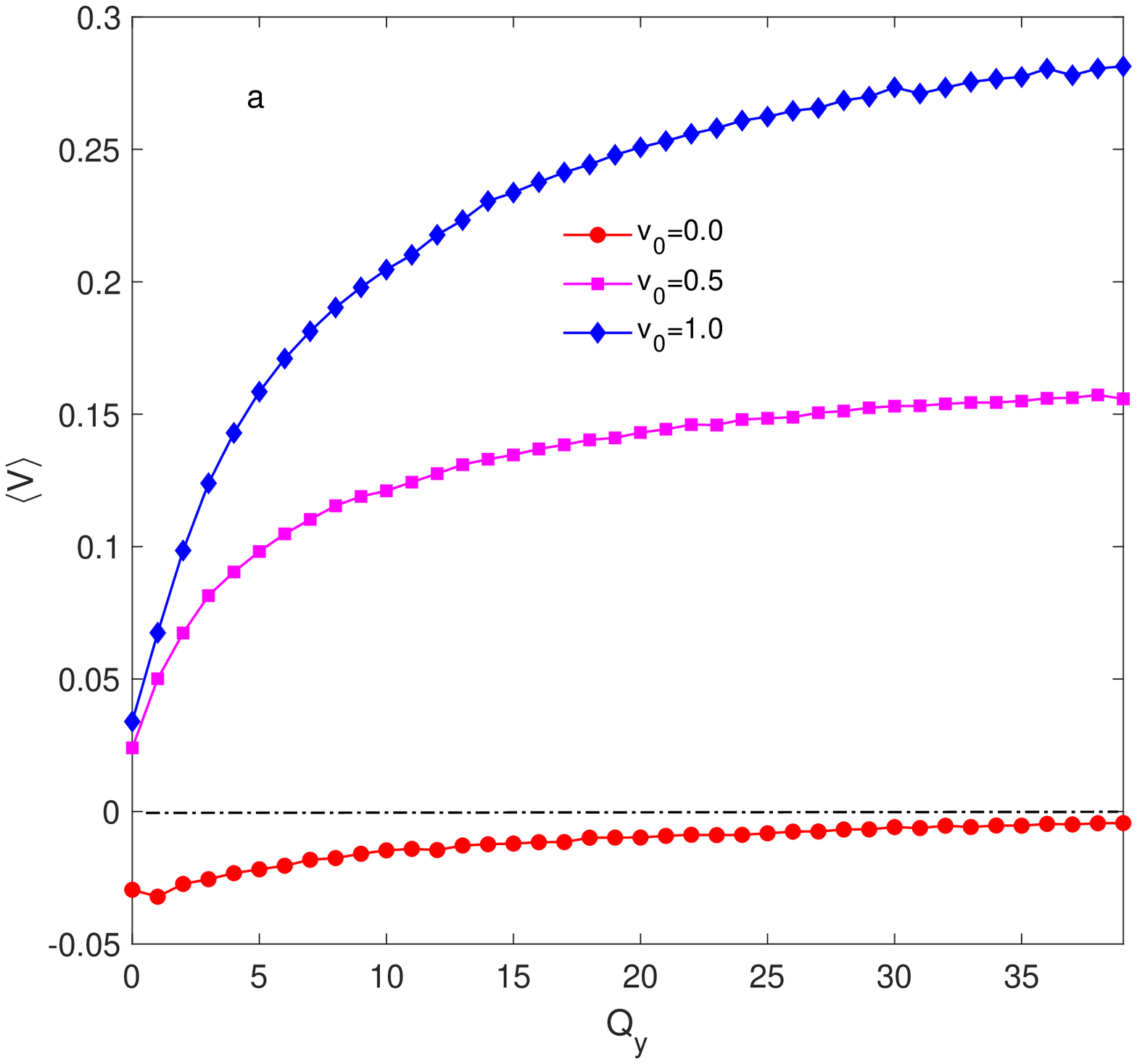}
}
\subfigure{
\includegraphics[height=6cm,width=6cm]{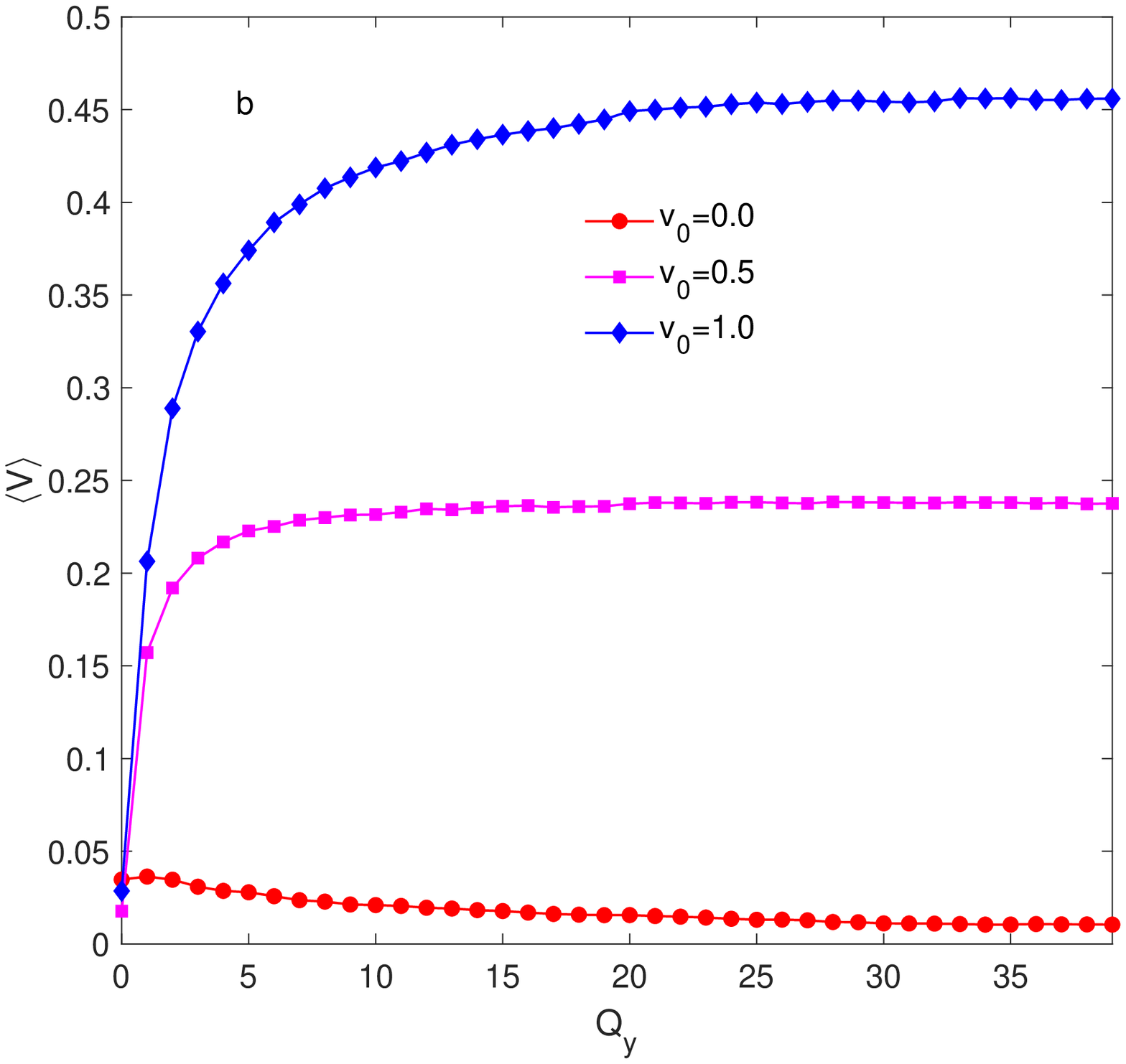}
}
\caption{The average velocity $\langle V\rangle$ as a function of $y$-axis noise intensity $Q_y$ with different $v_0$. The other parameters are $Q_x=2.0$, $Q_{\theta}=0.2$, $\tau_x=\tau_y=\tau_{\theta}=1.0$, $L=1.0$, $\bar{k}=1.0$, $f=0.5$:(a)$\Delta k=-0.5$, (b)$\Delta k=0.5$.}
\label{VQy}
\end{figure}

Fig.\ref{VQy} shows $\langle V\rangle$ as a function of the $y$-axis noise intensity $Q_y$ with different $v_0$. In Fig.\ref{VQy}(a)($\Delta k=-0.5$), when $v_0=0.5$ and $v_0=1.0$, $\langle V\rangle$ increases with increasing $Q_y$, so large $y$-axis noise intensity helps to the $+x$ directional transport. In Fig.\ref{VQy}(a), when $v_0=0$, we can also find $\langle V\rangle<0$ and $\langle V\rangle$ shows little change when $Q_y$ is large. In Fig.\ref{VQy}(b)($\Delta k=0.5$), we find $\langle V\rangle$ increases with increasing $Q_y$ when $v_0=0.5$ and $v_0=1.0$, and $\langle V\rangle$ decreases with increasing $Q_y$ when $v_0=0.0$. In Fig.\ref{VQy}(b), we also find $\langle V\rangle$ shows little change when $Q_y$ is large.

\begin{figure}
\centering
\subfigure{
\includegraphics[height=6cm,width=6cm]{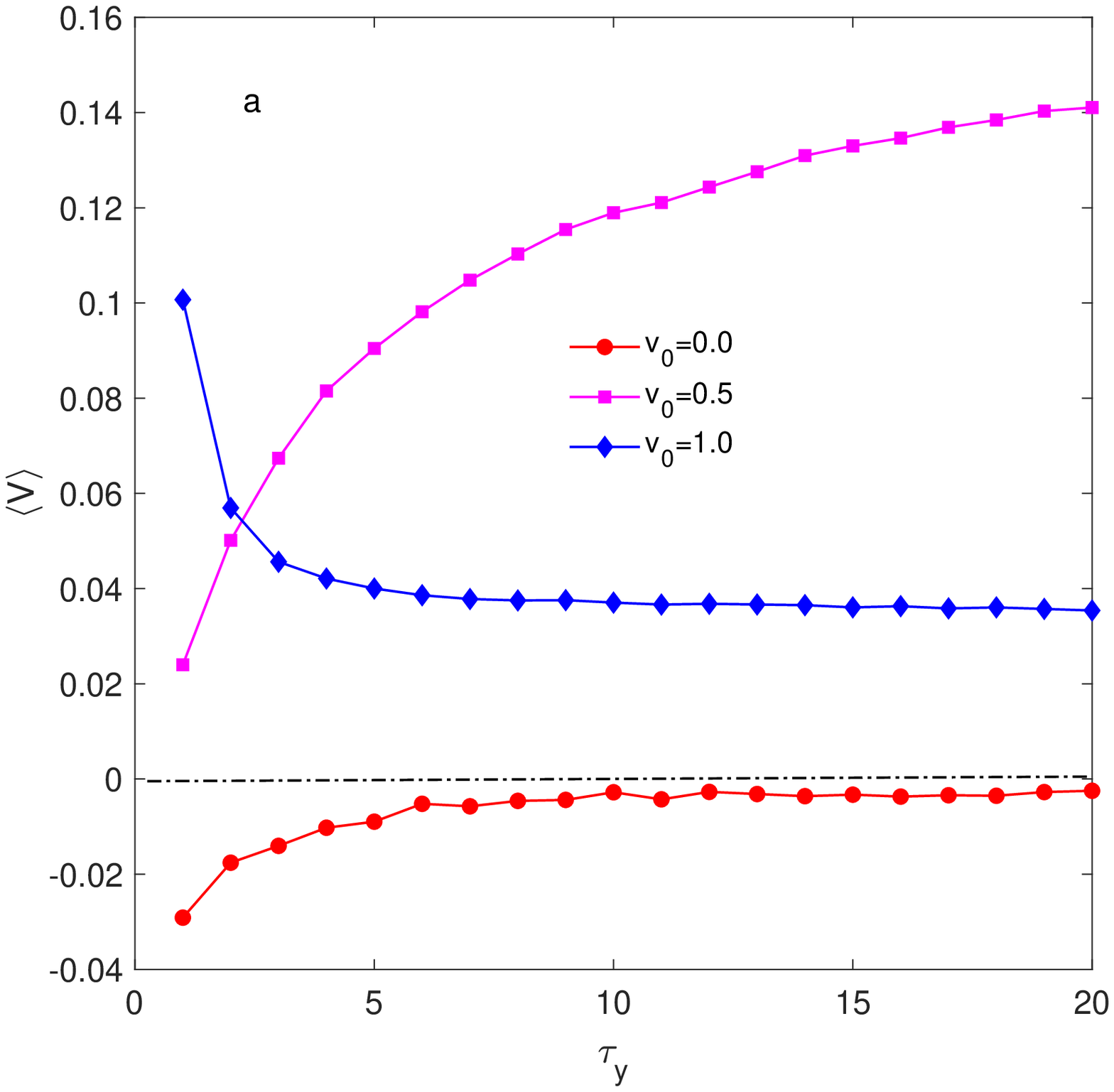}
}
\subfigure{
\includegraphics[height=6cm,width=6cm]{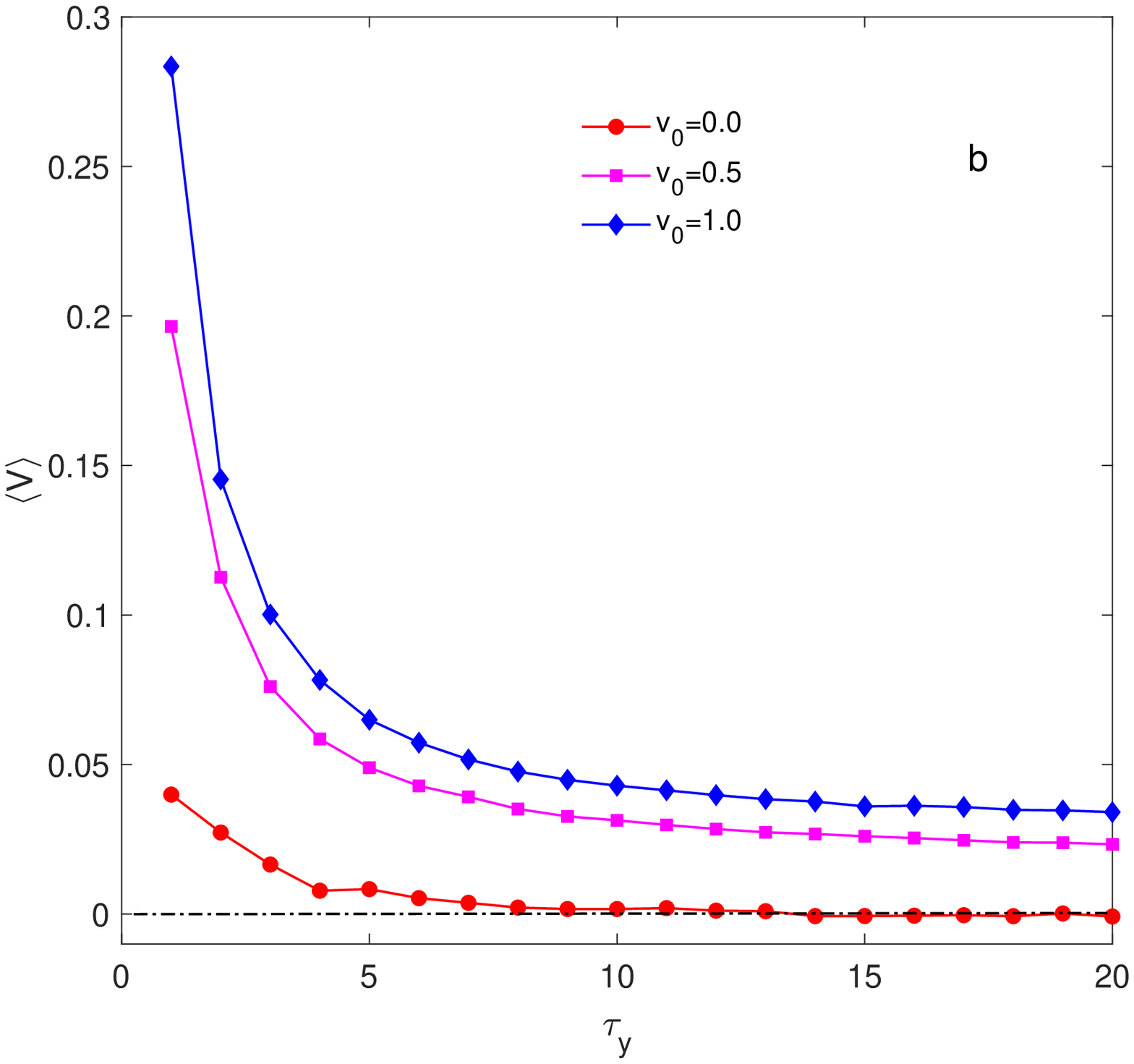}
}
\caption{The average velocity $\langle V\rangle$ as a function of $y$-axis noise self-correlation time $\tau_y$ with different $v_0$. The other parameters are $Q_x=Q_y=2.0$, $Q_{\theta}=0.2$, $\tau_x=\tau_{\theta}=1.0$, $L=1.0$, $\bar{k}=1.0$, $f=0.5$:(a)$\Delta=-0.5$, (b)$\Delta=0.5$.}
\label{Vtauy}
\end{figure}

Fig.\ref{Vtauy} shows the average velocity $\langle V\rangle$ as a function of the  $y$-axis noise self-correlation time $\tau_y$ with different $v_0$. In the case of $\Delta k=-0.5$(Fig.\ref{Vtauy}(a)), $\langle V\rangle$ increases with increasing $\tau_y$ when $v_0=0.0$ and $v_0=0.5$, and $\langle V\rangle$ decreases with increasing $\tau_y$ when $v_0=1.0$, and $\langle V\rangle$ shows little change when $\tau_y$ is large. In Fig.\ref{Vtauy}(a), we also find $\langle V\rangle>0$ when $v_0=0.5$ and $v_0=1.0$. When $v_0=0.0$, $\langle V\rangle<0$, and $|\langle V\rangle|$ decreases with increasing $\tau_y$. In the case of $\Delta k=0.5$(Fig.\ref{Vtauy}(b)), we find $\langle V\rangle>0$ and decreases with increasing $\tau_y$ when $v_0=0$($v_0=0.5$ and $v_0=1.0$).

\begin{figure}
\centering
\subfigure{
\includegraphics[height=6cm,width=6cm]{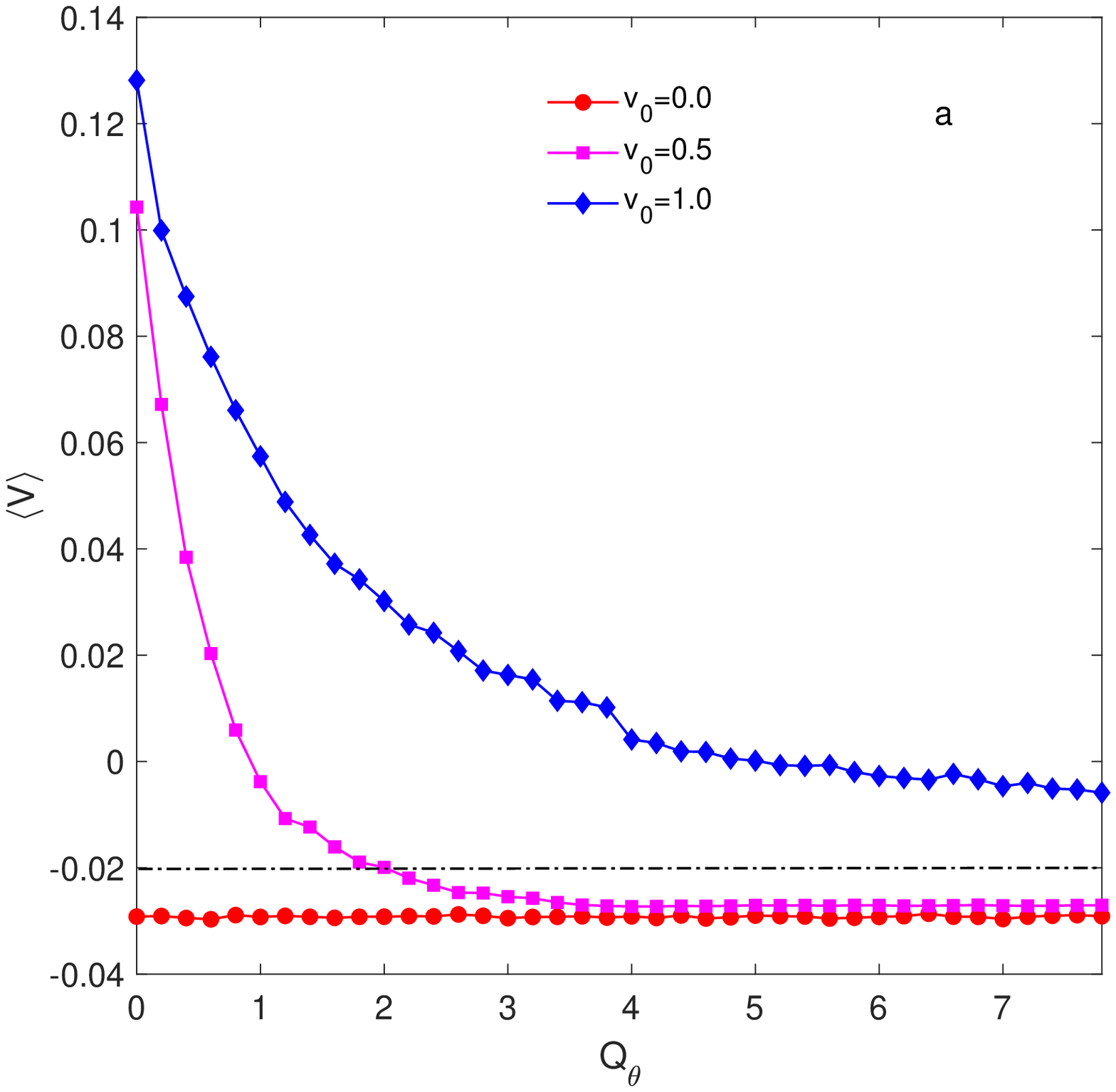}
}
\subfigure{
\includegraphics[height=6cm,width=6cm]{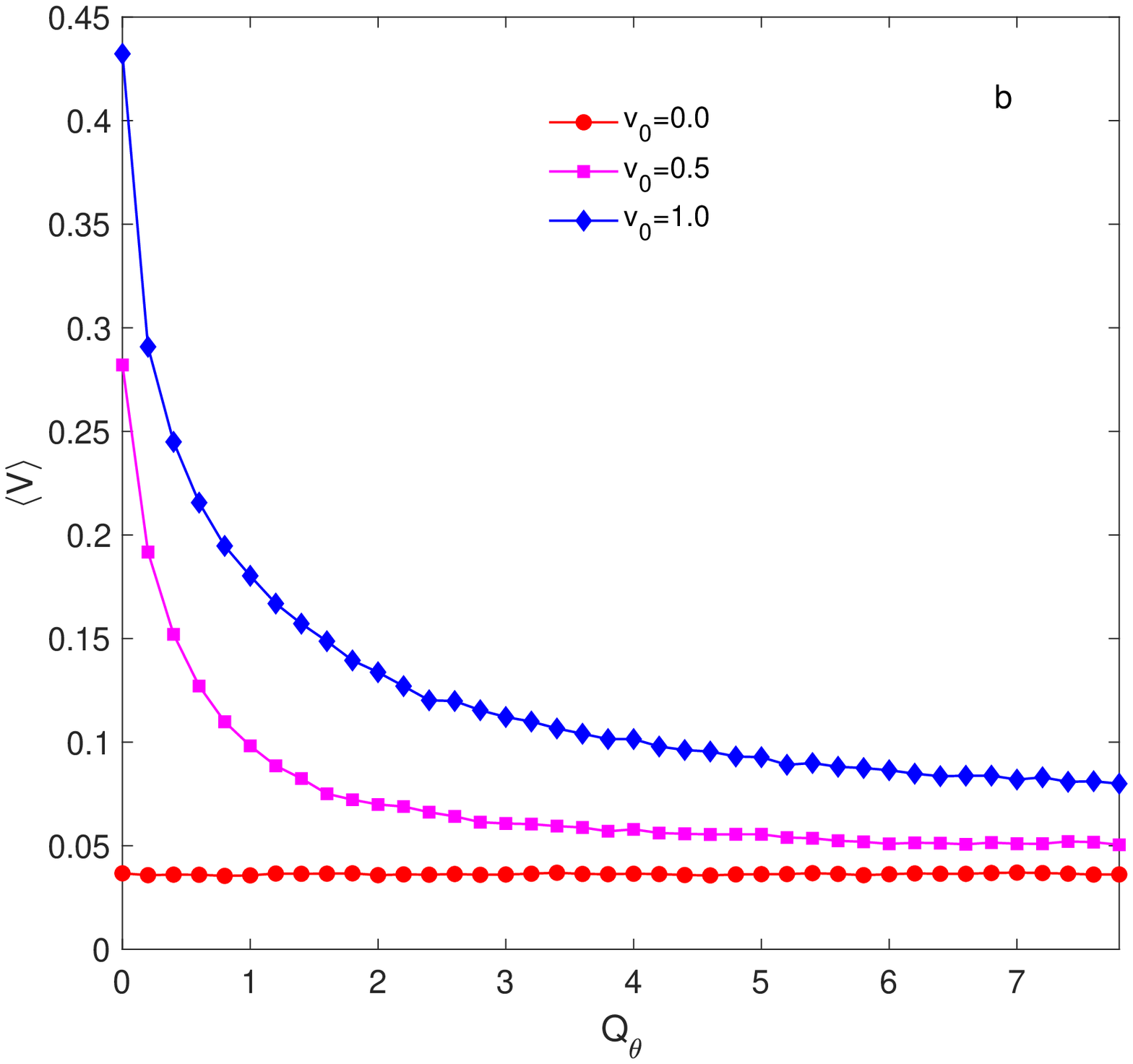}
}
\caption{The average velocity $\langle V\rangle$ as a function of angle noise intensity $Q_{\theta}$ with different $v_0$. The other parameters are $Q_x=Q_y=2.0$, $\tau_x=\tau_y=\tau_{\theta}=1.0$, $L=1.0$, $\bar{k}=1.0$, $f=0.5$:(a)$\Delta k=-0.5$, (b)$\Delta k=0.5$.}
\label{VQtheta}
\end{figure}

Fig.\ref{VQtheta} shows $\langle V\rangle$ as a function of angle noise intensity $Q_{\theta}$ with different $v_0$. In Fig.\ref{VQtheta}(a)($\Delta k=-0.5$), $\langle V\rangle$ decreases with increasing $Q_{\theta}$ when $v_0=0.5$ and $v_0=1.0$. In the case of $v_0=0.5$, $\langle V\rangle>0$ when $Q_{\theta}<2$, and $\langle V\rangle<0$ when $Q_{\theta}>2$, so the particle changes its moving direction with increasing $Q_{\theta}$. In Fig.\ref{VQtheta}(a), we also find that $\langle V\rangle$ always less than zero and changes very little with increasing $Q_{\theta}$ when $v_0=0.0$. In Fig.\ref{VQtheta}(b)($\Delta k=0.5$), $\langle V\rangle>0$ whenever $v_0=0.0$, $v_0=0.5$ and $v_0=1.0$. The particle move in $+x$ direction, and the moving speed decreases with increasing $Q_{\theta}$ when $v_0=0.5$ and $v_0=0.5$. In  Fig.\ref{VQtheta}(b), when $v_0=0.0$, $\langle V\rangle$ changes very little with increasing $Q_{\theta}$.

\begin{figure}
\centering
\subfigure{
\includegraphics[height=6cm,width=6cm]{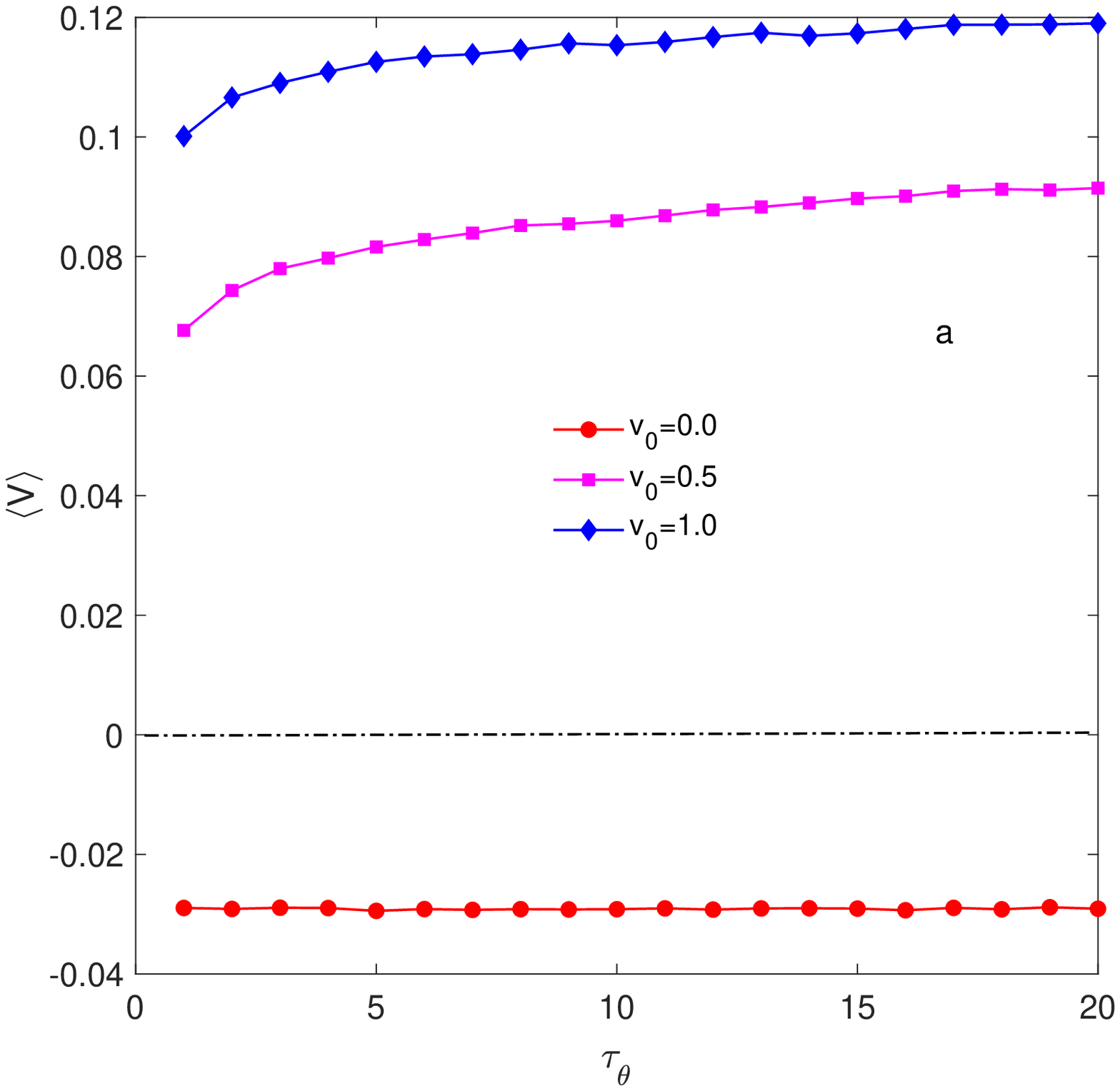}
}
\subfigure{
\includegraphics[height=6cm,width=6cm]{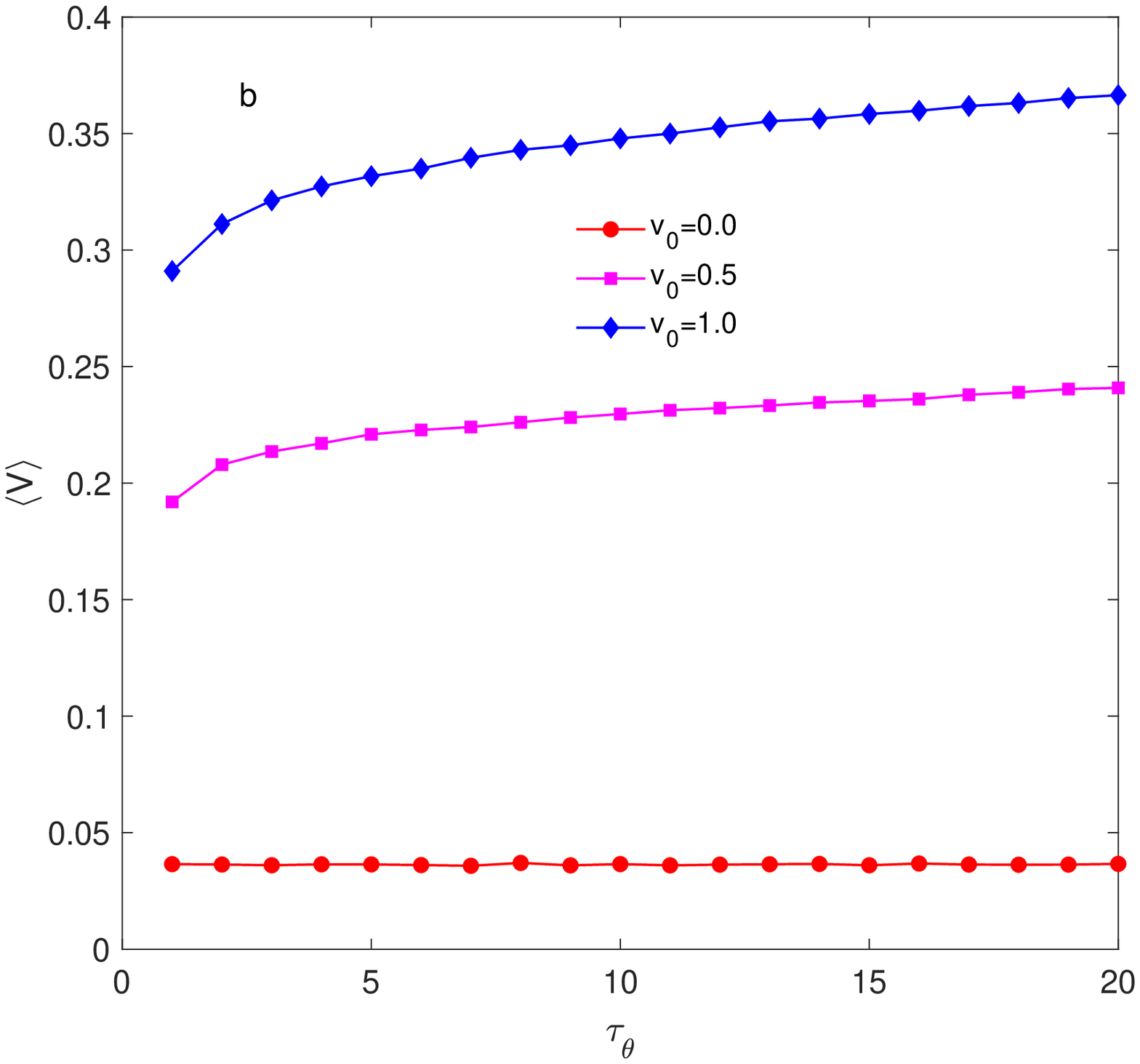}
}
\caption{The average velocity $\langle V\rangle$ as a function of angle noise self-correlation time $\tau_{\theta}$. The other parameters are $Q_x=Q_y=2.0$, $Q_{\theta}=0.2$, $\tau_x=\tau_y=1.0$, $L=1.0$, $\bar{k}=1.0$, $f=0.5$:(a)$\Delta k=-0.5$, (b)$\Delta k=0.5$.}
\label{Vtautheta}
\end{figure}

The average velocity $\langle V\rangle$ as a function of of angle noise self-correlation time $\tau_{\theta}$ with different $v_0$ is reported in Fig.\ref{Vtautheta}. In Fig.\ref{Vtautheta}(a)($\Delta k=-0.5$), $\langle V\rangle$ increases with increasing $\tau_{\theta}$ when $v_0=0.5$ and $v_0=1.0$, and $\langle V\rangle$ shows little change with increasing $\tau_{\theta}$ when $\tau_{\theta}$ is large. In Fig.\ref{Vtautheta}(a), when $v_0=0.0$, we find $\langle V\rangle<0$, and $\langle V\rangle$ changes very little with increasing $\tau_{\theta}$. In Fig.\ref{Vtautheta}(b)($\Delta k=0.5$), we find $\langle V\rangle>0$ when $v_0=0.0$($v_0=0.5$ and $v_0=1.0$), and $\langle V\rangle$ increases with increasing $\tau_{\theta}$ when $v_0=0.5$ and $v_0=1.0$. In Fig.\ref{Vtautheta}(b), when $v_0=0.0$,  $\langle V\rangle$ shows very little change with increasing $\tau_{\theta}$.

\begin{figure}
\centering
\subfigure{
\includegraphics[height=6cm,width=6cm]{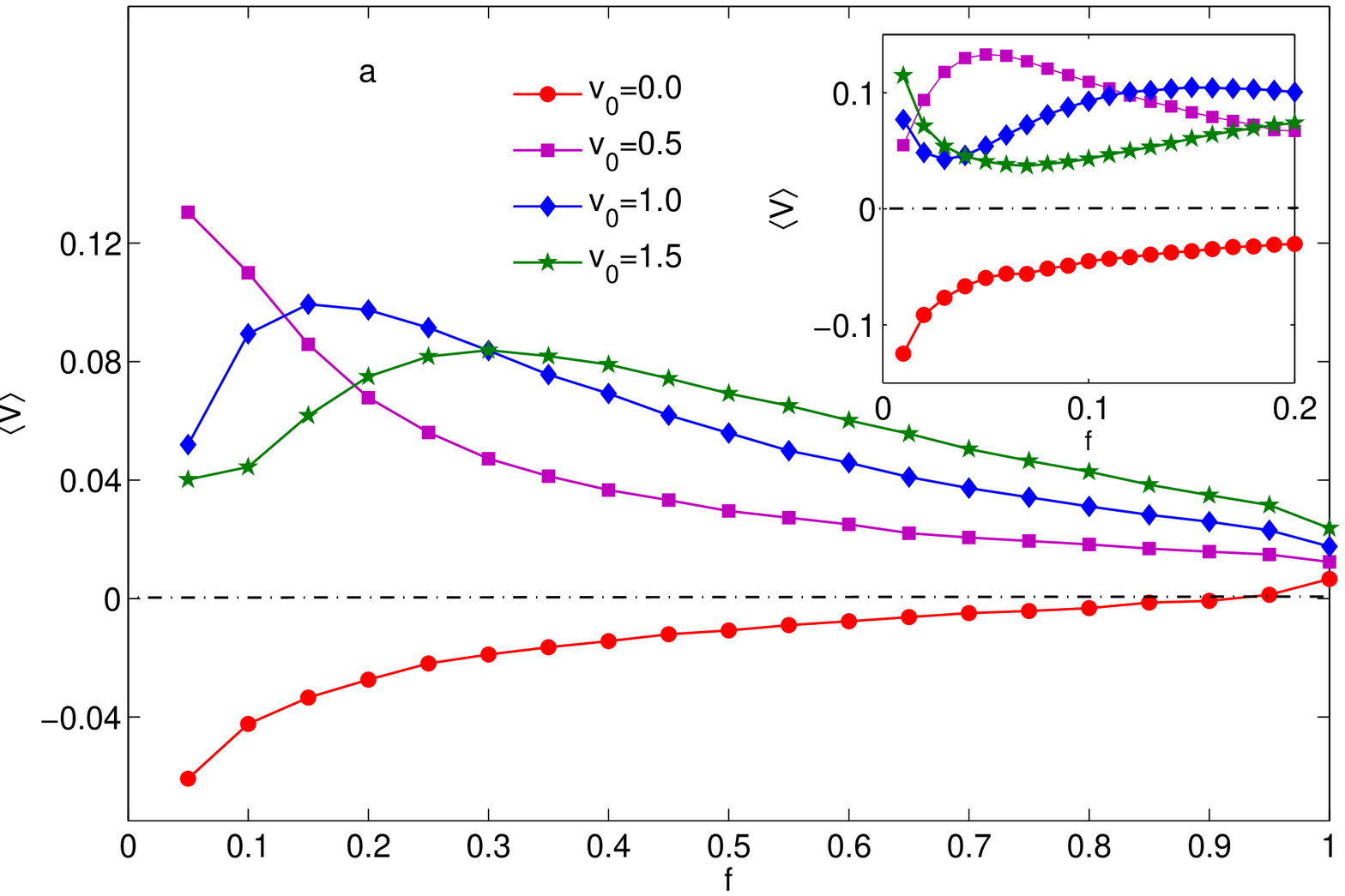}
}
\subfigure{
\includegraphics[height=6cm,width=6cm]{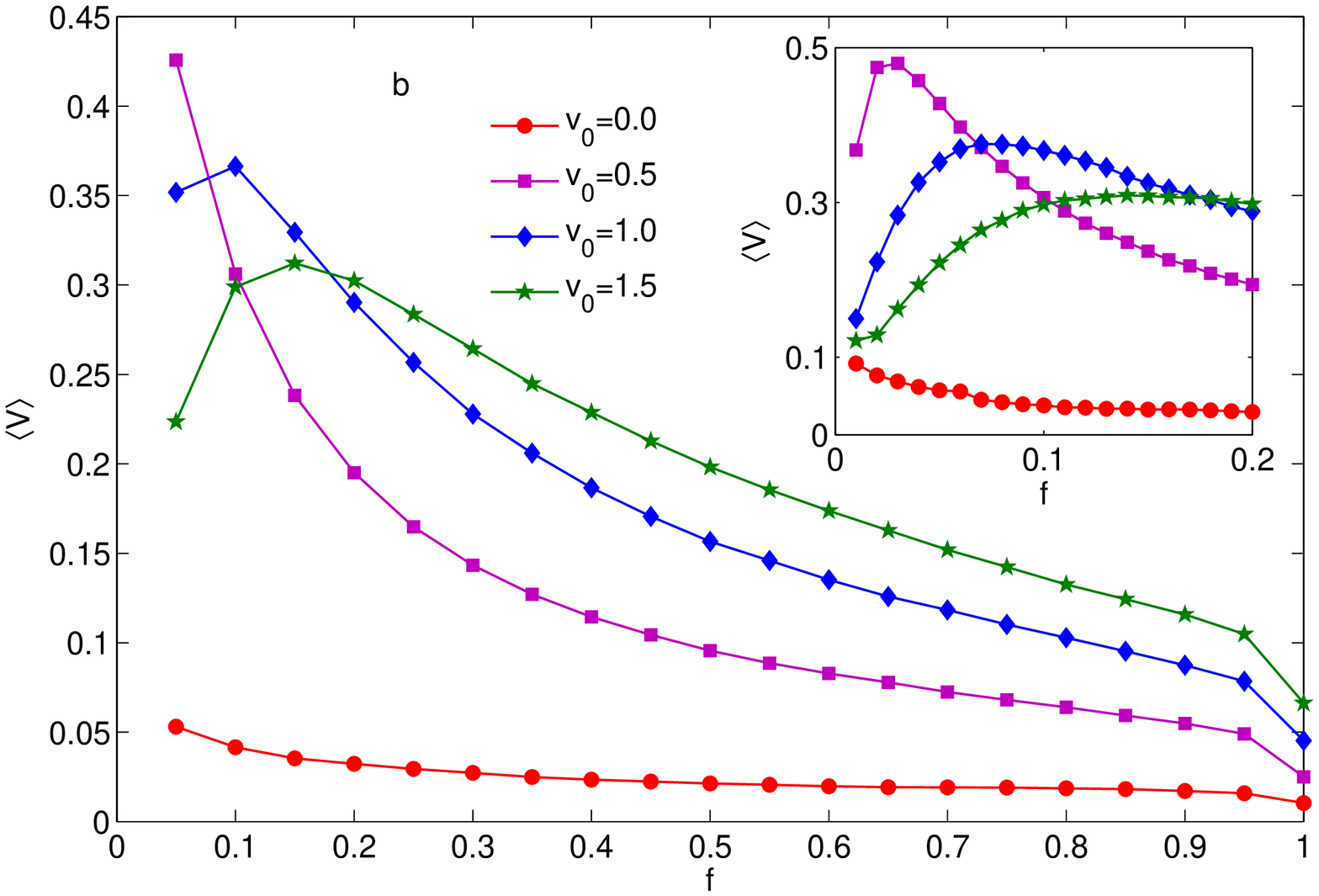}
}
\caption{The average velocity $\langle V\rangle$ as a function of the ratio $f$. The other parameters are $Q_x=Q_y=2.0$, $Q_{\theta}=0.2$, $\tau_x=\tau_y=\tau_{\theta}=1.0$, $L=1.0$, $\bar{k}=1.0$, $v_0=2.0$:(a)$\Delta k=-0.5$, (b)$\Delta k=0.5$.}
\label{Vxiaof}
\end{figure}

Fig.\ref{Vxiaof} shows the average velocity $\langle V\rangle$ as a function of $f$. In Fig. \ref{Vxiaof}(a)($\Delta k=-0.5$), when $v_0=0.0$,  $\langle V\rangle$ increases with increasing $f$($|\langle V\rangle|$ decreases with increasing $f$) and $\langle V\rangle<0$, so passive particle moves in $-x$ direction, and the moving speed decreases with increasing $f$. In Fig. \ref{Vxiaof}(a), when $v_0=0.5$, $\langle V\rangle>0$ and decreases with increasing $f$, so large value of $f$ is bad for directional transport. In Fig. \ref{Vxiaof}(a), when $v_0=1.0$ and $v_0=1.5$, we find $\langle V\rangle$ has a maximum with increasing $f$($\langle V\rangle_{max}=0.099$ at $f=0.15$ when $v_0=1.0$, and $\langle V\rangle_{max}=0.084$ at $f=0.3$ when $v_0=1.8$.), as $f$ is the ratio of the particle radius $R$ and the bottleneck half width $b$, so proper size of particle is good for transport, too large or too small size of particle has a negative influence on the transport. In Fig. \ref{Vxiaof}(b), $\langle V\rangle$ decreases with increasing $f$ when $v_0=0.0$ and $v_0=0.5$, so large $f$ is bad for directional transport. In Fig. \ref{Vxiaof}(b), $\langle V\rangle$ has a maximum with increasing $f$($\langle V\rangle_{max}=0.366$ at $f=0.1$ when $v_0=1.0$, and $\langle V\rangle_{max}=0.312$ at $f=0.15$ when $v_0=1.5$.), this means there exits an optimal value of $f$ at which the average velocity takes its maximal value.

\begin{figure}
\centering
\subfigure{
\includegraphics[height=8cm,width=10cm]{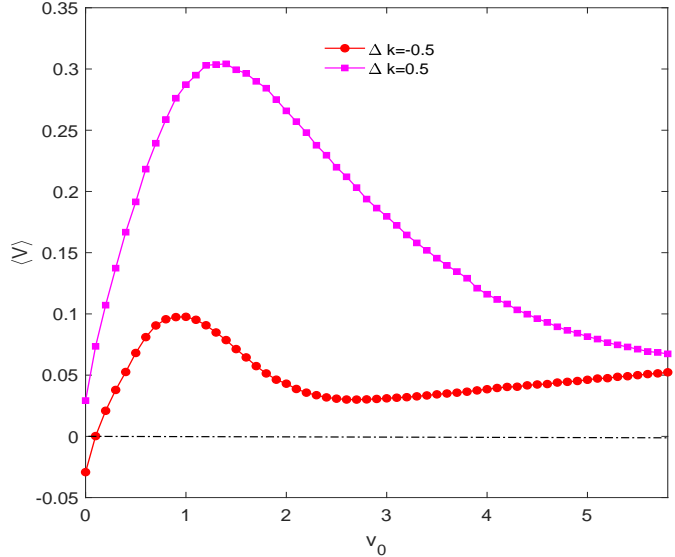}
}
\caption{The average velocity $\langle V\rangle$ as a function of $v_0$ with different $Delta k$. The other parameters are $Q_x=Q_y=2.0$, $Q_{\theta}=0.2$, $\tau_x=\tau_y=\tau_{\theta}=1.0$, $L=1.0$, $\bar{k}=1.0$, $f=0.5$.}
\label{Vv0}
\end{figure}

Fig. \ref{Vv0} shows $\langle V\rangle$ as a function of the self-propelled speed $v_0$ with different $\Delta k$. We find there exists an optimal value of $v_0$ at which $\langle V\rangle$ takes its maximum value, which means the appearance of resonance phenomenon. In the case of the $\Delta k=-0.5$, $\langle V\rangle\approx-0.02$ at $v_0=0$, so passive particle should moves in $-x$ direction. In the case of the $\Delta k=0.5$, $\langle V\rangle\approx0.02$ at $v_0=0$, passive particle moves in $x$ direction.

\begin{figure}
\center{
\includegraphics[height=8cm,width=10cm]{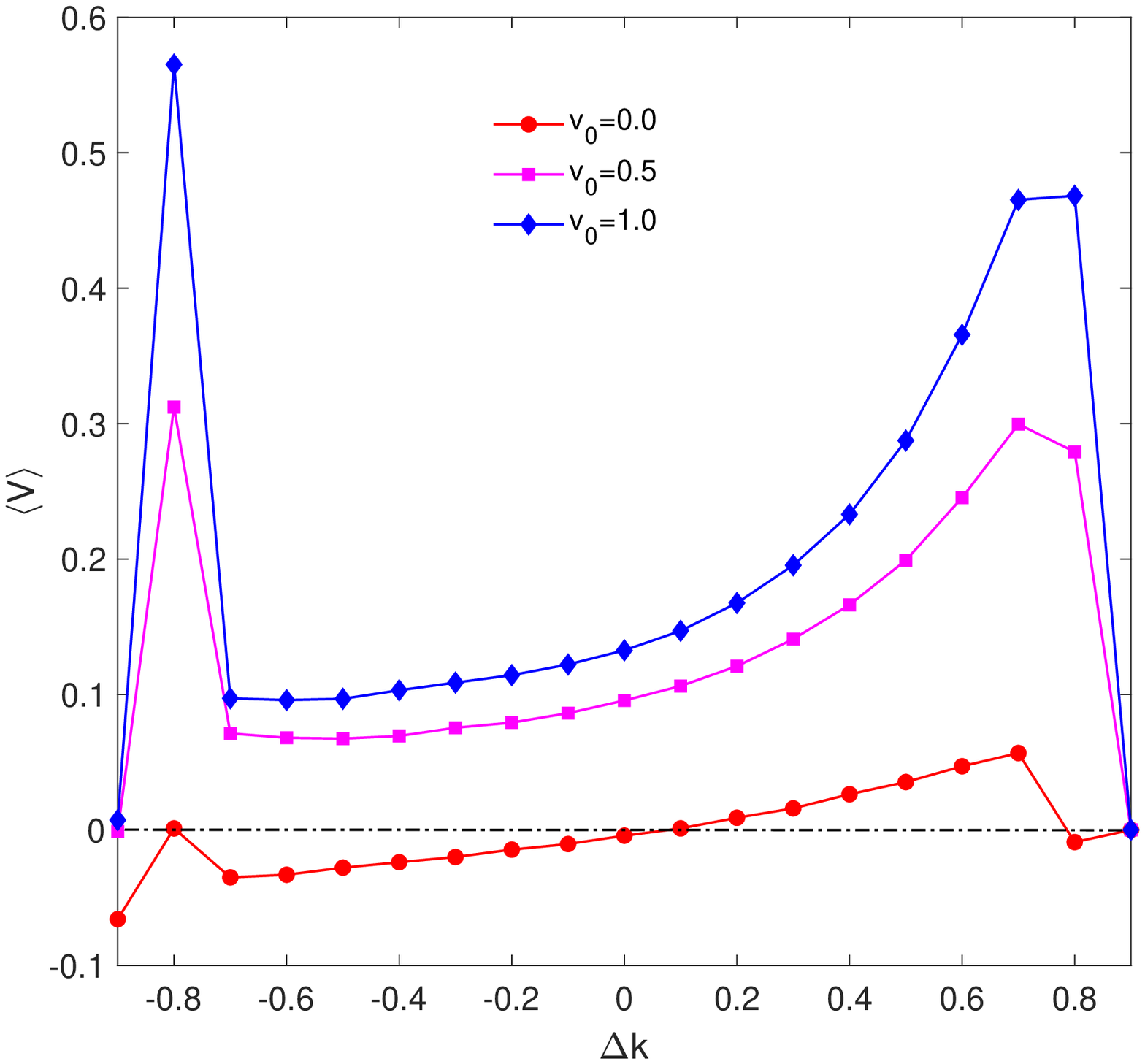}
\caption{The average velocity $\langle V\rangle$ as a function of $\Delta k$ with different self-propelled speed $v_0$. The other parameters are $Q_x=Q_y=2.0$, $Q_{\theta}=0.2$, $\tau_x=\tau_y=\tau_{\theta}=1.0$, $L=1.0$, $\bar{k}=1.0$, $f=0.5$.}
\label{VDeltak}
}
\end{figure}
The dependence of $\langle V\rangle$ on the asymmetry parameter $\Delta k$ with different $v_0$ is shown in Fig. \ref{VDeltak}. We find $\langle V\rangle$ exhibits complex behavior with increasing $\Delta k$. $\langle V\rangle$ has two maximums with increasing $\Delta k$. When $v_0=0.5$ and $v_0=1.0$, $\langle V\rangle\rightarrow0$ as $\Delta k=-0.9$, and reaches the first maximum quickly at $\Delta k=-0.8$, and then reduces to small value at $\Delta k=-0.7$, and then increases with increasing $\Delta k$ when $-0.7\leq\Delta k\leq0.7$, and reaches the second maximum at $\Delta k\approx0.7$, and finally reduces to approximately zero at $\Delta k=0.9$. For $v_0=0.0$, there exit two maximums in the $\langle V\rangle-\Delta k$ curve. $\langle V\rangle\approx-0.066$ when $\Delta k=-0.9$, which means the particle moves in $-x$ direction. $\langle V\rangle$ reaches the first maximum$\langle V\rangle\approx0$ at $\Delta k=-0.8$, and then reduce to small value($\langle V\rangle<0$) at $\Delta k=-0.7$, and then the particle changes moving direction from $-x$ to $x$ with increasing $\Delta k$ when $-0.7\leq\Delta k\leq0.7$.

\section{\label{label4}Conclusions}
In this paper, we numerically studied the transport phenomenon of self-propelled particle confined in a zigzag $2D$ channel with colored noise. The noise(noise parallel to $x$-axis and $y$-axis), the asymmetry parameter and the self-propelled speed have joint effect on the particle. We find the average speed in $x$ direction maybe increases or reduces with increasing $x$-axis noise intensity. Large $y$ axis noise intensity will strengthen $+x$ directional movement for self-propelled particles, but will restrain $-x$(or $+x$) directional movement for passive particle. $\langle V\rangle$ decreases monotonically with increasing angle noise intensity $Q_{\theta}$ for self-propelled particles, but have small effect on passive particle. In some cases too large or too small size of the particle is unfavorable for directional movement. The average velocity exhibits complicated behavior with increasing self-propelled speed.

\section{Acknowledgments}

Project supported by Natural Science Foundation of Anhui Province(Grant No:1408085QA11) and College Physics Teaching Team of Anhui Province(Grant No:2019jxtd046).

\bibliographystyle{elsarticle-num}

\end{document}